\documentclass[journal]{IEEEtran}

\usepackage{graphicx}
\graphicspath{ {Images/} }
\usepackage{array}
\usepackage[usenames, dvipsnames]{color}
\usepackage{contour}
\usepackage{verbatim}
\usepackage{blindtext}
\usepackage{float}
\usepackage{calc}
\def\BibTeX{{\rm B\kern-.05em{\sc i\kern-.025em b}\kern-.08em    T\kern-.1667em\lower.7ex\hbox{E}\kern-.125emX}}
    
\title{Super resolution imaging through the human skull}

\author{\IEEEauthorblockN{Danai E. Soulioti, David Esp\'indola, Paul A. Dayton and Gianmarco Pinton}  \\
\thanks{This work was supported by grants from the National Institute of Health R01-CA220681, R01-EB025419.}

\thanks{The authors are with the Joint Department of Biomedical Engineering, University of North Carolina at Chapel Hill and the North Carolina State University, Chapel Hill, North Carolina, USA.}
\thanks{This work has been submitted to the IEEE Transactions on Ultrasonics, Ferroelectrics, and Frequency Control for possible publication. Copyright may be transferred without notice, after which this version may no longer be accessible.}
\thanks{Corresponding author: G. Pinton }
\thanks{Contact: gfp@unc.edu, soulioti@email.unc.edu}
}

\begin{document}

\maketitle

\begin{abstract}
High resolution transcranial ultrasound imaging in humans has been a persistent challenge for ultrasound due to the imaging degradation effects from aberration and reverberation. These mechanisms depend strongly on skull morphology and they have high variability across individuals. Here we demonstrate the feasibility of human transcranial super-resolution imaging using a geometrical focusing approach to concentrate energy at the region of interest, and a phase correction focusing approach that takes the skull morphology into account. It is shown that using the proposed focused method, we can image a 208$\mu$m microtube behind a human skull phantom in both an out-of-plane and an in-plane configuration. Individual phase correction profiles for the temporal region of the human skull were calculated and applied to transmit-receive a custom-focused super-resolution imaging sequence through a human skull phantom, targeting the microtube, at 68.5mm in depth, at 2.5 MHz. Microbubble contrast agents were diluted to a concentration of 1.6$\times$10$^6$ bubbles/mL and perfused through the microtube. It is shown that by correcting for the skull aberration, the RF signal amplitude from the tube improved by a factor of 1.6 in the out-of-plane focused emission case. The lateral registration error of the tube's position, which in the uncorrected case was 990 $\mu$m, was reduced to 50$\mu$m in the corrected case as measured in the B-mode images. Sensitivity in microbubble detection for the phase corrected case increased by a factor of 1.5 in the out-of-plane imaging case, while in the in-plane case it improved by a factor of 1.3 while achieving an axial registration correction from an initial 1885$\mu$m error for the uncorrected emission, to a 284$\mu$m error for the corrected counterpart. These findings suggest that super-resolution may be used more generally as a clinical imaging modality in the brain.
\end{abstract}


\section{Introduction} \label{intro}
The ability to detect and image small vessels in the human brain is crucial especially in the context of neurodegenerative disorders such as Alzheimer's disease~\cite{alzheimer} and vascular dementia~\cite{EFSNguide}, as well as the assessment of stroke-related pathologies \cite{ctstroke}. To appropriately identify the presence of cortical or subcortical infarcts or other stroke lesions using imaging methods, high spatial resolution is required. \cite{Gorelick2672}. Three dimensional Computed Tomography Angiography (CTA), one of the gold-standard imaging methods of brain vessels, can reliably resolve structures down to 0.4~mm \cite{3dct}. Magnetic Resonance Angiography (MRA) offers comparable resolution with an effective voxel size of 0.39~mm \cite{MRA}, while two dimensional Digital Subtraction Angiography (DSA) offers reconstruction down to a pixel size of 0.34~mm \cite{DSA}. With venous vessel sizes in the brain measuring as small as a few micrometers \cite{vessel}, the importance of increasing the resolution of standard imaging methods is evident. 

An invaluable tool that has shown great promise in the field of vascular imaging, is ultrasound Contrast Enhanced Super Resolution (CESR) imaging, also known as Ultrasound Localization Microscopy (ULM)~\cite{ULM}. CESR has gained tremendous popularity during recent years, and has been used to generate microvascular maps that surpass the ultrasound diffraction limit, 
and can potentially produce images resolving vascular structures with a diameter as small as 9~$\mu$m, thus providing a $\lambda/10$ resolution limit~\cite{EC2015}. 

Inspired by the use of fluorescent ``blinking targets'' in optical super resolution methods~\cite{SRhuang}, CESR in ultrasound uses gas-filled encapsulated microbubbles with a mean diameter on the order of a few microns as contrast agents that are injected in the blood flow. These microbubbles  stochastically appear and disappear in the field of view. At sufficiently low concentrations, the appearing microbubbles can be spatiotemporally separated, allowing for their localization at a sub-wavelength level~\cite{FL2017}. 

One of the challenges of CESR is the isolation of the microbubbles from the surrounding tissue in the background. To achieve microbubble separation, various filtering methods have been employed, such as high-pass spatiotemporal filters which can distinguish fast moving components from  slow ones ~\cite{EC2015}. Singular value decomposition (SVD) filters~\cite{SVD} can also be used for the separation by exploiting the variation in the coherence length between microbubbles and tissue, which is especially useful when the bubbles and tissue are moving  at the same velocity, i.e. in the case of a  frequency overlap. Following the filtering process,  microbubbles are localized with centroid detection techniques~\cite{acousticMB}, which are robust and computationally efficient. Alternatively, correlation analysis is another option that has been used successfully in optical super-resolution~\cite{correlation} and it has the advantage of multiple iterations to converge to a solution, which is especially useful in very aberrating and high-clutter scenarios, where the signals can benefot from iterative isolation. 

In 2015, Errico {\it et al}. ~\cite{EC2015}, proposed the use of microbubble contrast agents at a concentration of $2\times10^8$ bubbles/mL, allowing for the generation of a super resolved image within a clinically reasonable duration (150~s). Using this method, they produced super-resolved images of a rat brain with a maximum spatial resolution of 8$\mu$m. Recent studies have also verified the method's {\it in vivo} feasibility. Namely, Viessmann {\it et al} imaged the vasculature of the mouse ear~\cite{CJK2015} with a maximum resolution of 20~$\mu$m laterally and Lin {\it et al} (\cite{FL2016,FL2017}) produced super resolved images of tumor angiogenesis with vessel sizes  ranging from 5 to 150~$\mu$m.

Although its capabilities far surpass the resolution limits of conventional ultrasound, super-resolution imaging is not devoid of limitations, especially from a clinical perspective. In recent {\it in vivo} studies, for example, the depth penetration of the method was limited to approximately 2~cm ~\cite{CJK2015,FL2017}. This is mainly the result of the drop-off in bubble detectability with imaging depth, which can be attributed to several mechanisms, including attenuation in the soft tissue, reverberation, multiple scattering effects and off-axis clutter that significantly degrade the images. Compared to conventional imaging, these effects are compounded by the low amplitude of the emissions, which is used to avoid microbubble destruction. This limitation has been addressed, for example, in previous work, where a simultaneous multi-focus approach combined with high frame-rate acquisitions was used to target multiple deep targets~\cite{adaptive}, making the super-resolution of 150~$\mu$m microtubes at a depth greater than 6~cm feasible. 

Another challenge lies in the fact that to satisfy the separability condition, which is a prerequisite for bubble isolation and localization, a low concentration of microbubbles is used. When combined with the low flow rate in vessels (especially capillary vessels which have a flow rate in the order of 2~$\mu$L/min ~\cite{flow}), this results in the need for the acquisition of a considerable number of frames (on the order of thousands) to fully populate a vessel~\cite{FL2016,OMA2013}.
This subsequently results in prolonged acquisition times which poses a challenge not only for patient time management, but also for motion correction,  which attempts to reduce artifacts caused by breathing or heart cycles~\cite{artifact}. 

In the context of using ultrasound in the human brain, in imaging and in therapy, the biggest challenge is ultrasonic propagation through the skull. The heterogeneous nature of the human skull bone at a microscopic level, causes considerable distortion of the ultrasonic beam, in the form of resolution compromising reverberations, refraction~\cite{refraction}, multiple scattering, reflection between bone and tissue interfaces and mode conversion wth the three latter being the main sources of attenuation~\cite{procdegradGP}, while absorption plays a secondary role at the higher frequencies used in imaging \cite{GPatten}. More specifically, the skull thickness, density and equivalent sound velocity tend to vary across the length of the transducer causing phase aberration of the wavefront. The combination of these factors results in a decrease in amplitude, as well as a three dimensional shift in the position of the target in the resulting image.

To overcome this persistent challenge, several phase aberration correction techniques, based on time reversal, speckle brightness, pitch-catch methods (\cite{fink96,tanter98,aberration1,aberration2,aberration3}), have been developed. A large number of phase correction techniques have been developed for transcranial focused ultrasound therapy. Focused ultrasound can thermally or mechanically destroy malignant brain tissue. This topic has been investigated for the last 60 years and has seen a resurgence in the last 2 decades, due to its non-invasive nature and its potential for localized ablation and efficient penetration \cite{braintherapy}. 
Correction techniques have been successfully applied to correct the focal position accuracy and intensity on emission in transcranial ultrasound therapy, with registration error corrections with respect to the treatment location ranging from 0.5~mm to 2.3~mm \cite{Almquist2016,cttherapy}. Such techniques have also been shown to substantially improve image quality and target resolution through the skull both in real-time and in conventional imaging~\cite{phase_smith}. 

It should be noted that in therapy, a one-way correction is necessary for the improvement of target focusing, whereas in imaging the correction must be calculated and applied twice, on both the emission and on the reception. This makes the process more sensitive to the quality of the correction. Moreover, when compared to ultrasound imaging, ultrasound therapy is performed at lower frequencies, at which the skull behaves in a less challenging way when compared to a higher frequencies. This is due to several mechanisms, including the fact that the trabecular structure size distribution is comparable to the wavelengths employed in higher-frequency imaging (100-150~$\mu$m)~\cite{trabec}, that lower frequencies are less attenuated, and that lower frequencies are less sensitive to aberration from the skull morphology.

The difficulties present in transcranial imaging have stalled its growth compared to therapy, and other imaging methods are usually preferred to depict the brain vasculature at high resolution. Most significant work in transcranial brain imaging has been performed using Transcranial Doppler ultrasound (TCD), a robust method in vascular imaging and flow measurements in the brain that has been used for more than 50 years \cite{kato65}. Due to its ability to record cerebral hemodynamic changes in real-time, TCD is often paired with cognitive studies \cite{dopplercog}. Color Doppler imaging (CDI) can demonstrate the relative direction and velocity of blood flow in brain vessels, superimposed on a conventional B-mode ultrasound image of the stationary tissue. Doppler imaging is subject to the same limitations as conventional ultrasound as far as image quality is concerned, namely significant registration errors and insufficient spatial resolution for detection of sub-wavelength targets. To address these limitations, groups have employed 3D helmet-like array structures with 3D Doppler capabilities in conjunction with phase corrections \cite{dopplerBL} which can generate images of millimeter sized vascular structures, but they are still not capable of imaging smaller vessel anomalies and capillary structures, which require sub-wavelength resolution.

One of the pioneer studies in the field of super-resolution transcranial imaging, showed that a 255~$\mu$m radially sized microtube can be successfully super-resolved transcranially using a 2D therapy array at 612~kHz~\cite{OMA2013}. However, even with a total accumulation of 3000 frames, it is challenging to fully populate the tube with detected bubbles, leaving some uncertainty in the exact shape and size of the tube.  

Here, we demonstrate the feasibility of super-resolution imaging of 208~$\mu$m targets through an {\it ex vivo} human skull, by performing experiments using focused ultrasound and high frame-rate focused sequences on a clinical linear ultrasound transducer array. Two experimental configurations are presented, imaging a microtube placed in the out-of-plane direction and in the in-plane direction with respect to the transducer, both with a diameter of 208~$\mu$m, at a depth of 68.5~mm. Phase corrections that take into account the individual skull morphology are applied to improve the sensitivity in bubble detection and the accuracy of target registration in both lateral and axial dimensions. Micro-Computed Tomography derived maps of a human skull are converted into acoustical properties, and then used as as input to a custom simulation tool (called Fullwave) that we have developed to model acoustic propagation through the skull. Simulations of acoustic propagation through the skull are used to determine the expected maximum theoretical improvements for phase corrected imaging of sub-resolution targets. 

\begin{figure}[ht]
\setlength{\unitlength}{0.5\textwidth}
\begin{picture}(0.8,0.6)(0,0)
\put(0.0,0.0){\includegraphics[trim= 0 0 10 10,clip,height=0.6\unitlength]{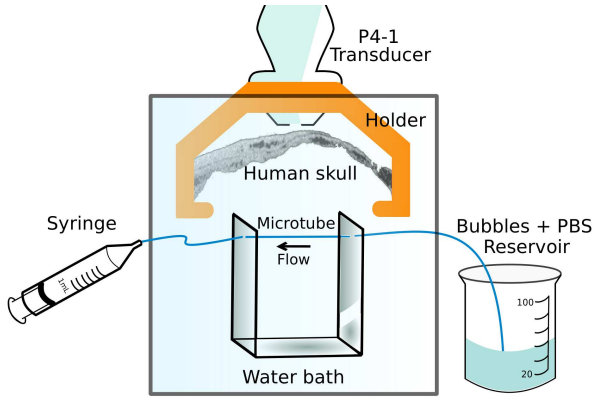}}
\end{picture}
\caption{Schematic of the experimental setup. A custom 3D printed holder secured the skull in place while a separate tube holder allowed for both tube configurations with 90 degree rotations.}
\label{fig:setup}
\end{figure}

\section{Methods} \label{method}

A cleaned human skull specimen was cut in half across the sagittal suture. The temporal region of the skull was used for imaging since it is relatively thin, and therefore provides a good acoustical window. To remove air trapped in the porous structure of the skull, it was degassed for 12 hours in water using a vacuum pump. A sketch of the experimental setup can be seen in Fig.~\ref{fig:setup}. A custom designed skull and transducer holder was 3D printed and used to rigidly secure the skull in place with respect to the transducer. Since the phase correction is sensitive to movement, this holder ensured that the skull remained immobile throughout the duration of the whole experiment and also allowed for the attachment and detachment of the skull for different acquisitions, hence ensuring the stability and repeatability of the experiment. In addition, an adjustable microtube holder was built to secure a thin-walled polycarbonate microtube, with an inner diameter of 208~$\mu$m and an outer diameter of 250~$\mu$m (Paradigm Optics Inc., WA, USA), in place. The holder can rotate by 90 degrees around the axial dimension, which allowed the microtube to move from an out-of-plane configuration to an in-plane one. These parts were placed in a 56~L tank filled with degassed water.

Lipid-encapsulated microbubble contrast agents containing decafluorobutane were formulated as described in previous work~\cite{LINbubbles}, and were diluted to a concentration of $1.6\times10^6$~bubbles/mL. These microbubbles were measured with an Accusizer (Accusizer 780, PALS-Particle Sizing Systems, FL, USA) and typically have a mean diameter of less than 1~$\mu$m. A pump driven syringe was attached to one end of the tube to induce constant flow of the microbubble contrast agent solution. The syringe was refilled through a continuously mixed reservoir of microbubble solution at a refill rate of 75~$\mu$L/min. This configuration allowed for the establishment of uninterrupted, continuous flow, as well as the capability of replacing the reservoir when needed. Since bubbles can burst and aggregate in the bottom of the container, to ensure that the concentration remains constant throughout the experiment, the bubble reservoir was mixed as well as replaced, and flow was re-established, after every 6 minutes.

Imaging was performed with a Verasonics Vantage research scanner (Verasonics Inc., Redmond, WA, USA) and a P4-1 phased array transducer that transmitting 1 cycle pulses at 2.5~MHz and at a framerate of 700~Hz. A focused wave with a focal depth set at the position of the microtube was emitted. A more detailed description of the imaging sequence is given in subsection ~\ref{out}.

The two experiments, with the microtube in two different orientations with respect to the transducer, as well as the extraction of the correction are described in detail in subsections \ref{phase},\ref{out} and \ref{inplane}.

\subsection{Phase profile extraction for focal correction} \label{phase}

Prior to mounting the skull, a microtube filled with air, oriented in the out-of-plane direction with respect to the imaging field, was aligned at the center of the transducer, laterally, and at 6.85~cm of depth. A single full-aperture focused emission, with the parameters mentioned above and with an amplitude of 168~kPa, targeted the microtube position. The whole aperture was also used to receive the RF data. 

This data served as a reference while computing the skull-distorted version of the tube profile and the relative lag between the two. As expected, the backscattered reflection from the microtube appears as a parabolic phase profile. The phase associated with the microtube was isolated and flattened by subtracting the equivalent delay of the focused emission, which was also a parabola. In principle the emitted focused profile should be identical to the backscaterred phase profile. However, this was not exactly true, due to minor alignment errors in the positioning of the microtube. These alignment errors can be attributed to the manual alignment of the microtube. 
A cross-correlation algorithm, which correlates data from each channel of the transducer to a reference channel of the same data set, was applied to extract the microtube profile.

\begin{figure}[ht]
\setlength{\unitlength}{0.5\textwidth}
\begin{picture}(0.8,0.5)(0,0)
\put(0.0,0.0){\includegraphics[trim= 25 0 35 10,clip,height=0.48\unitlength]{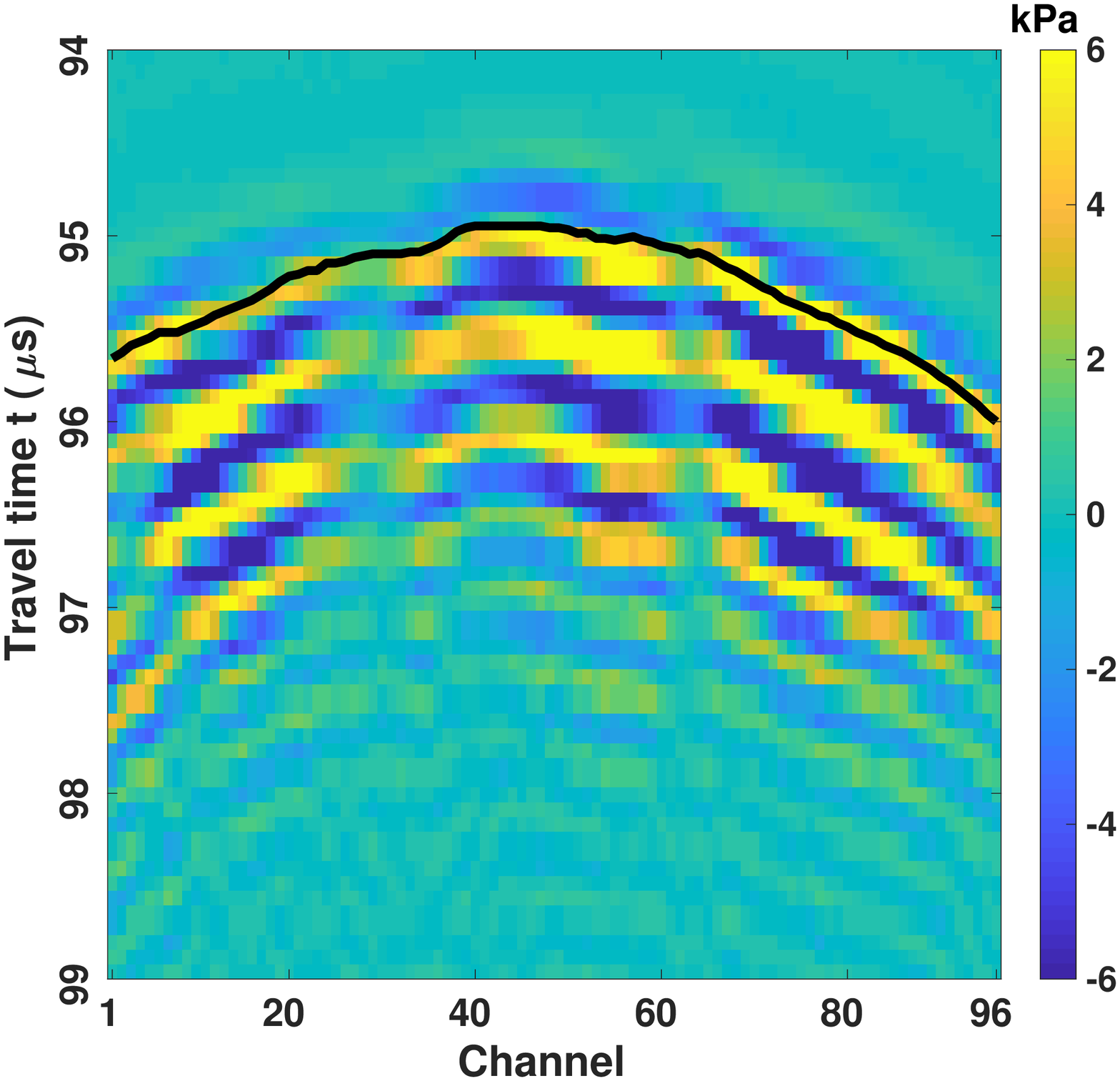}}
\put(0.52,0.0){\includegraphics[trim= 32 0 20 10,clip,height=0.47\unitlength]{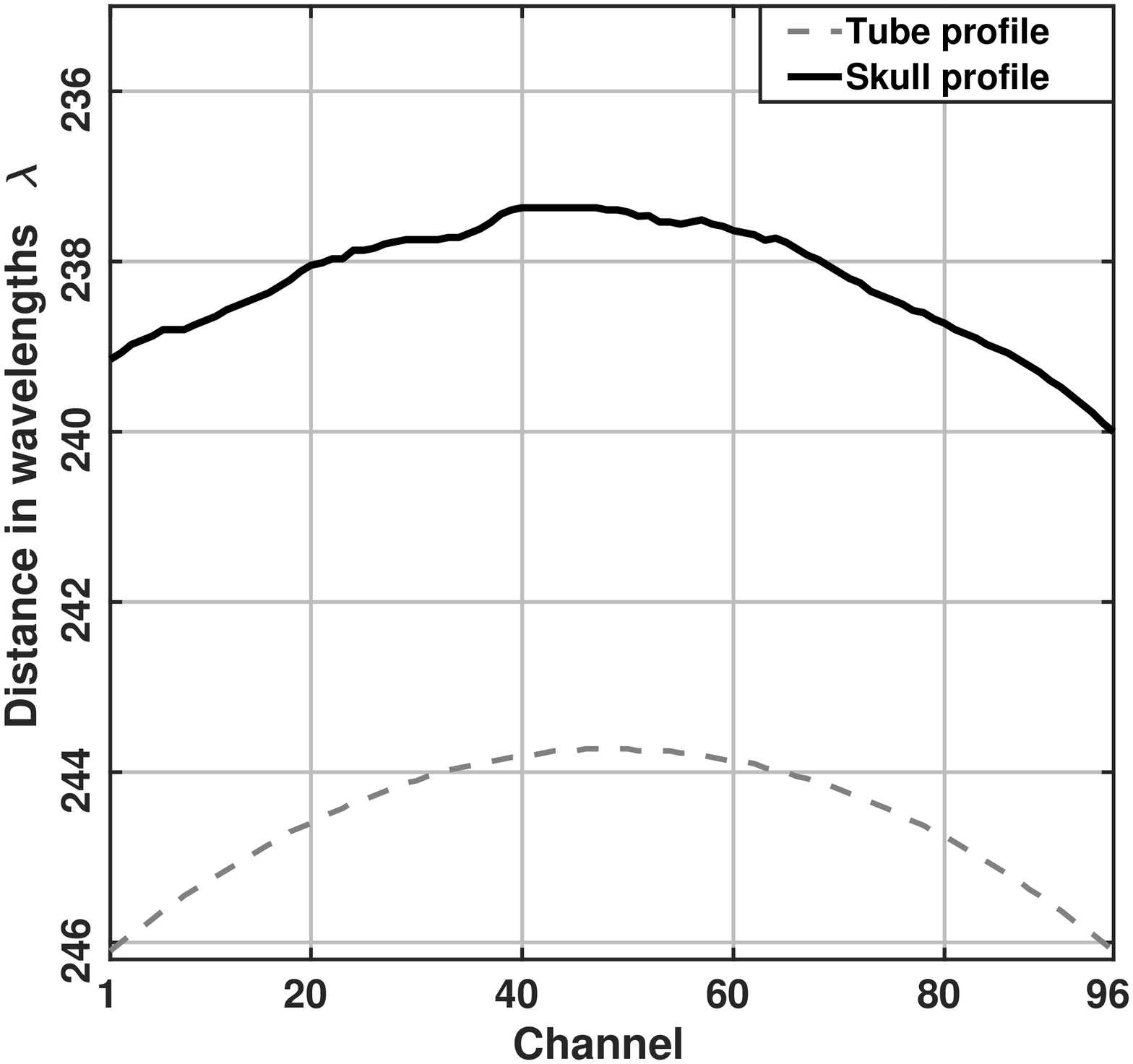}}
\put(0.38,0.068){\small A}
\put(0.93,0.068){\small B}
\end{picture}
\caption{(A) The detected profile of the microtube underneath the skull measured with the correlation detection method (black line), plotted on the RF data. (B) Detected profile of the target without the skull and the target through the skull in time units expressed as wavelengths. The RMS difference between the two profiles was 90.3~ns.} 
\label{fig:rf}
\end{figure}

Subsequently, without perturbing the position of the tube and transducer, the skull was mounted on the special holder and stabilized so that it remained in a fixed registered position  throughout the experiment. A second acquisition, using the same single focus sequence, was performed at the same target, but at 2.21~MPa to account for the attenuation from the skull and to deliver approximately the same pressure amplitudes at the microtube location. The raw data acquired through the skull shown in Fig.~\ref{fig:rf} (A) corresponds to the backscattered signal from the microtube, after it has crossed the skull twice, on emission and reception. The interaction of the pulse with the skull causes variations in amplitude arising from interference in parts of the phase profile, as well as a distortion of the shape, which is no longer spherical. The same process, of flattening the image and correlating to a reference channel, was repeated as described above, and yielded the skull-distorted phase profile. The extracted skull-distorted profile is plotted on top of the raw data as a black solid line.

The tube profile and the skull-distorted tube profile, can be seen in Fig.~\ref{fig:rf} (B). An average offset of 6.32~$\lambda$ was measured between the two profiles. This offset is caused by the significant difference in the speed of sound between the skull and the medium. These profiles are also notably different in shape, which is a direct result of the aberration caused by the skull. The root mean square value for the difference in time between these two profiles, can also serve as a quantitative measure of skull induced aberration. In this particular acoustic window, aberration in time was estimated to be 90.3~ns (0.07$\lambda$). The skull-distorted profile served as input to the phase corrected version of the sequence that was implemented on the Vantage system to apply the correction on emission, and later the relative difference between the two profiles was used to correct on reception. The final B-mode images were thus corrected both ways.

\subsection{Out-of-plane experiment} \label{out}
Once the profiles were extracted, microbubble flow was established in the microtube. Three acquisitions were performed: First, the microtube only, second, a skull-microtube experiment using an uncorrected conventional emission and lastly, a skull-microtube experiment with the focusing correction on emission and reception. To apply the correction on emission, the previously extracted skull-distorted profile, was inverted in time, and was used as a delay input in the single focus sequence described in subsection~\ref{phase}.

For each of the three acquisitions, 1000 frames were recorded. Note that, in the absence of the skull a 168~kPa emission was used whereas when the skull is mounted a 1.35~MPa emission was used.  While the skull is absent, lower pressures must be applied to ensure that the bubbles will not burst, whereas with the attenuating skull present, higher pressures can be applied, still within a mechanical index lower than 1.5. Note that the amplitude is lower while the microbubble flow is being imaged, as compared to the 2.21~MPa used in subsection~\ref{phase} for the air-filled microtube imaging, to further ensure the integrity of the microbubbles, but at the same time provide sufficient energy.  

To correct on reception, each channel of the raw data from the corrected emission acquired, was delayed accordingly to the relative lag between the tube profile and the skull-distorted profile. This yields data that is corrected both ways. Following this step, all three datasets, were beamformed using a $\lambda$/12 spaced grid and a conventional delay-and-sum algorithm. 

An SVD filter was applied to the beamformed ultrasound images to isolate bubble appearances from the static background~\cite{SVD}, by removing the highest 2 singular values. A centroid detection algorithm was used to localize the microbubbles in the filtered images on a 40~$\mu$m grid. A bubble size filter was used to only detect bright spots that have a size larger than one wavelength, which is the minimum size of a point spread function. The accumulation of the position of all the detected bubbles among the 1000 acquired frames lead to the final super resolution image. 

\subsection{In-plane imaging} \label{inplane}
The microtube was subsequently placed in-plane with respect to the imaging plane. Three different cases were studied; no skull (the microtube alone), an uncorrected conventional emission through the skull and a corrected emission through the skull.
To scan the entire length of the tube, the full aperture was steered along different angles. The imaging field was divided into 96 different locations spanning a distance of 30~mm, spaced at a constant interval of $\lambda$/2. The steering angle was calculated based on the grid spacing and ranged from -0.2 to 0.2 radians. The focal depth was kept constant at 6.85~cm for all focusing locations. The whole-aperture was also used to receive 1000 frames at each focal position. This resulted in a full field scan comprising 96 different acquisitions and a cumulative total of 96000 frames. A full lateral scan had a duration of 6~min. For the tube without the skull 168~kPa were emitted and for the cases where the skull is present, 1.35~MPa were emitted.

For the corrected emission through the skull, the correction extracted as described in section~\ref{phase}, was added to each steered delay profile for each focus. The same correction profile was overlaid on the steering profiles even though the correction is local in nature, i.e. it is extracted from a target in the center of the transducer imaging field. This is consistent with the assumption that the relative distance between the skull and the transducer is smaller than the distance between the skull and the target. Therefore the spatial dependence of the skull distortion is assumed to be small. A more sophisticated approach would use different correction profiles for different imaging positions. However it was expected that a single phase correction would correct the registration errors in some neighborhood of the central target. 

For the correction on receive, each of the 96 acquisitions was corrected using the relative lag between the tube and skull-distorted profiles prior to beamforming, as mentioned in section~~\ref{out}. Beamforming of the RF data was performed using a conventional delay-and-sum algorithm that took into account the steering angle. Each of the 96 sets of a 1000 frames was beamformed independently, and was subsequently filtered with the SVD filter which removed the 3 highest singular values for the microtube only case, and the 2 highest for both the experiments through skull. Subsequently, bubble appearance events for each of the 96 sets were individually localized using the centroid detection algorithm across the whole imaging field. This resulted in 96 separate full-field super resolution images. This is due to the fact that although the foci are localized, microbubbles are detected up to 1~mm away from the target position. The 96 individually generated super resolution images were combined, by arithmetic summation, to produce the final super-resolved image depicting the whole microtube.

\subsection{Simulations} \label{sims}
The Fullwave Simulation tool is a numerical solver based on finite differences in the time domain (FDTD)~\cite{pinton2009fullwave} which can be used to model ultrasound propagation with a high dynamic range. This tool has been extensively used for various acoustic applications. A few examples include simulating the thermal effects of focused ultrasound in transcranial brain therapy~\cite{fullwave2}, modeling of acoustic cavitation risk in the brain~\cite{fullwave1} and transcranial focused ultrasound parameter optimization~\cite{fullwave3}. 

Here this numerical tool was used to produce two dimensional simulations in which the experimental conditions were closely mimicked.  The sources of image degradation in ultrasound imaging depend not only  on aberration of phase profile clutter, which are corrected, but also on multiple scattering or reverberation artifacts, which are not corrected~\cite{pintondegrad,pintontrans}. Simulations also provide a more controlled acoustical environment, where the properties of interest can be investigated without the interference of the experimental setup. The aim of the simulations is to determine the imaging improvements under a best-case phase aberration correction scenario and allow us to investigate the one-way distortion of the ultrasonic beam, which can in turn provide us with a more effective means to phase correcting.

A CT scan of the same skull specimen used in the experiment was converted to speed of sound, attenuation and density maps that were used in the simulations. A 2D field of view from the temporal region of the skull was selected. The maps of the simulation were calibrated with amplitude measurements at the focus to match the experimental pressure conditions. The density and speed of sound maps were calculated as linearly scaled versions of the CT image in Hounsfield units with maximum values of 1850~$kg/m^3$ and 2900 $m/s$, respectively~\cite{duck,aubryHIFU}. This map conversion process was has been previously used and validated for focused ultrasound surgery and has been used for transcranial simulations with the Fullwave tool~\cite{pintontrans,aubryHIFU}. The size of the simulation field was 80~mm in depth and 32~mm in width. Transmitted pulses had the same parameters as described in the experimental setup. The grid size was set to 38.5~$\mu$m which is equivalent to 16 points per wavelength and a Courant-Friedrichs-Lewy~\cite{cfl} condition of 0.3 was used. A target, with a single pixel size, was placed at 6.85~cm in depth. The transducer was located at the top of the medium and the maximum pressure of 2.21~MPa at the emission was matched to hydrophone measurements for the P4-1 transducer.

The simulated acquisition process was the same as the experimental process described previously in subsection~\ref{phase}. Three acquisitions were performed, one of the microtube without the skull, an uncorrected emission through the skull, and lastly a corrected emission through the skull. To derive the correction profile, the target was used as a point source, i.e. the experimentally described pulse was emitted directly from the target, propagating the wave through the skull, which and was then received by the transducer above the skull (one way travel). In this manner, the wave crosses the skull only once, hence undergoes half the distortion. The received raw data was processed as described in subsection ~\ref{phase}. By means of correlation, a skull-distorted profile was directly detected. A second emission was performed by adding this profile to the focused delay emitted from the transducer, which constituted the corrected emission. The received data were first corrected on reception, following the same process of imposing a relative lag to each channel, and then beamformed as previously described in section~\ref{method}.

\begin{figure}[ht]
\setlength{\unitlength}{0.5\textwidth}
\begin{picture}(0.9,0.85)(0,0)
\put(0.0,0.0){\includegraphics[trim= 230 0 235 0,clip,height=0.85\unitlength]{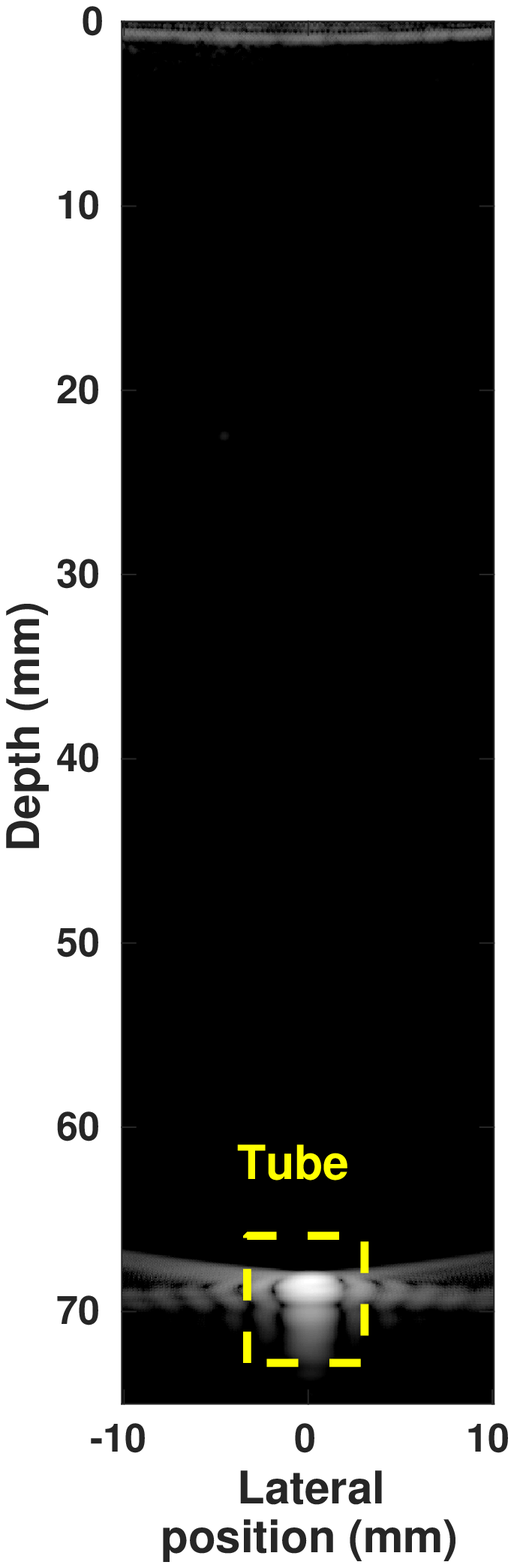}}
\put(0.315,0.0){\includegraphics[trim=270 0 235 0,clip,height=0.85\unitlength]{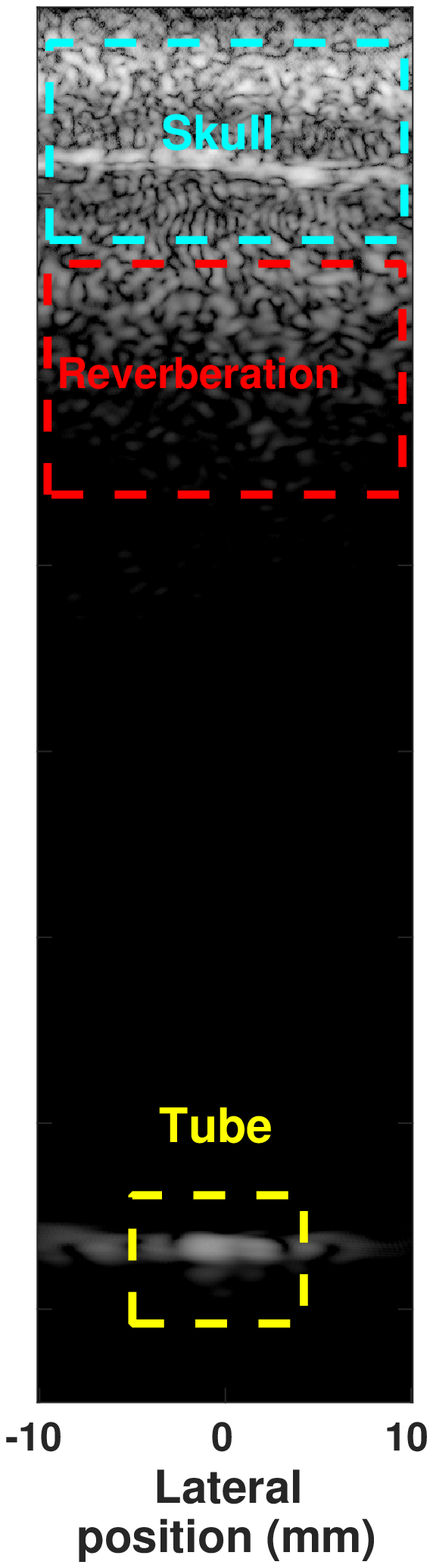}}
\put(0.59,0.0){\includegraphics[trim=250 0 200 0,clip,height=0.85\unitlength]{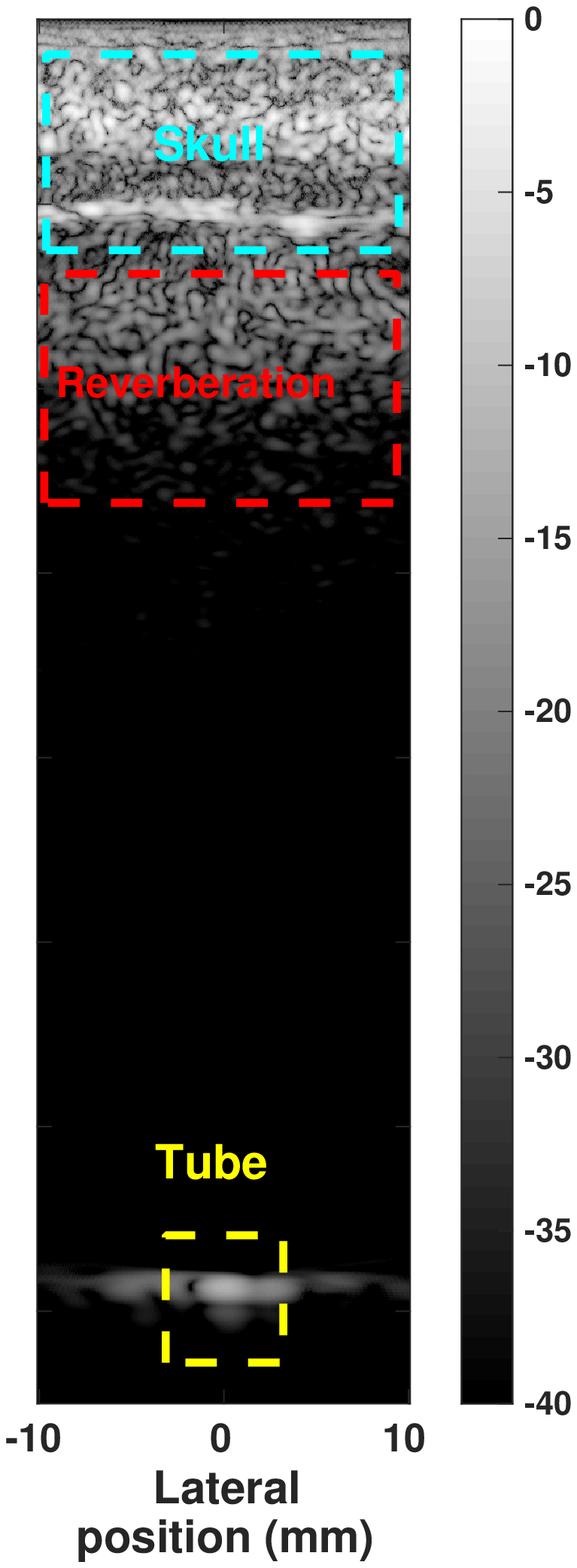}}
\color{white}
\put(0.25,0.1){\small A}
\put(0.51,0.1){\small B}
\put(0.765,0.1){\small C}
\end{picture}
\caption{B-mode images images for (A) the tube without skull, (B) uncorrected and (C) corrected emissions through skull. Scale for all images is in dB. The amplitude improves by a factor of 1.6 in the corrected case.} \label{fig:bmodes}
\end{figure}

\section{Results} \label{results}
\subsection{Out-of-plane imaging and simulations} \label{results_out}
The B-mode images from the out-of-plane target experiment, shown in Fig.~\ref{fig:bmodes}, can be used to characterize the conventional delay-and-sum point spread function (PSF) of the imaging system. The reference PSF of the microtube without the skull is shown in (A) and the B-mode image generated by the conventional uncorrected focused emission is shown in (B). The skull causes the point target to shift both laterally, as a result of aberration, and to move shallower, possibly as a result of aberration and definitely due to the increase in the average speed of sound. The shape of the PSF in (B) also appears to be degraded when compared to the microtube only B-mode image in (A). In the B-mode image generated by the corrected emission (Fig.~\ref{fig:bmodes}C), the maximum signal amplitude compared to the uncorrected B-mode increases by a factor of 1.6 (3.9~dB), defined as the ratio of amplitudes $I_c/I_u$ where $c$ stands for the corrected and $u$ for the uncorrected emission. In this image, the position appears to have been corrected and the shape also qualitatively improved. In both uncorrected and corrected B-modes (Fig. \ref{fig:bmodes}B and C), the skull is visible in the first 10~mm of depth, followed by a reverberation zone spanning approximately 20~mm. Note that no significant reduction of reverberation due to phase correcting is observed. 

\begin{figure}[ht]
\setlength{\unitlength}{0.5\textwidth}
\begin{picture}(0.5,0.5)(0,0)
\put(0.0,0.0){\includegraphics[trim= 8 0 0 5,clip,height=0.5\unitlength]{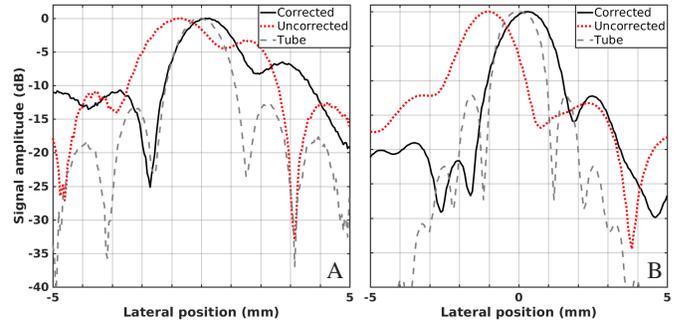}}
\put(0.46,0.07){\small A}
\put(0.93,0.07){\small B}
\end{picture}
\caption{(A) Beamplots for all three experimental cases, the lateral registration error of the tube’s position in the uncorrected case was 990~$\mu$m and in the corrected case it was 50~$\mu$m, as measured in the point target B-mode images. (B) Simulation beamplots for a different position of the skull in the temporal region. Correction was derived from a point source in the place of the target.} 
\label{fig:beamplot}
\end{figure}

To quantify the improvement produced by the phase correction, beamplots are plotted for each emission case Fig.~\ref{fig:beamplot}. The beamplots were calculated by averaging across all the frames and extracting the profile as a function of lateral position at the depth where the target exhibited its maximum amplitude, since there is an offset in depth between the B-modes for each acquisition and a global depth cannot be determined.  
Fig.~\ref{fig:beamplot}A shows the experimental beamplots derived from the three B-modes, corrected and uncorrected through the skull and the microtube only, on a normalized dB scale. The difference in position of the maximum of the main lobe between the corrected, uncorrected cases and the microtube (ground truth) yields the lateral registration error $\Delta$L.
The registration error laterally, was measured as 50~$\mu$m for the corrected case and at 990~$\mu$m for the uncorrected case yielding an improvement of 940~$\mu$m. Registration errors were also estimated axially, referred to as $\Delta$D, to be 300~$\mu$m and 1790~$\mu$m  respectively, exhibiting a 1490~$\mu$m improvement. The side lobes for the tube profile in Fig.~\ref{fig:beamplot} (A), are not symmetrical and deviate slightly from a classical point target shape due to minor alignment issues, and due to the fact that the microtube was manually aligned using live imaging, therefore there was a human error component present. Furthermore, since a solution of microbubbles was being imaged, the exact position of the microbubble can deviate from the actual center of the transducer and center of the microtube. The uncorrected profile is substantially degraded in shape, with asymmetrical side lobes that are hard to discern. The side lobe amplitude is 10.5~dB for the left side lobe (LSL) and 3.7~dB for the right side lobe (RSL). After applying the correction, the shape, especially of the main lobe, appears to be partially restored, however the side lobes remain high. Specifically, the left side lobe has an amplitude of 9~dB whereas the right of 7~dB. At -5~dB, the width of the main lobe (MLW) of the corrected profile is 522~$\mu$m larger than the width of the microtube only profile, whereas the width of the uncorrected profile is 1921~$\mu$m larger, due to the degradation in shape of its right side lobe.

\begin{table*}[t]
\centering
\begin{scriptsize}
   \caption{Summary comparison for the out-of-plane experiment.}
  \begin{tabular}{|c|c|c|c|c|c|c|c|c|c|c|c|c|c|c|c|c|}
    \hline
     &
      \multicolumn{6}{|c|}{\bf{Experimental B-mode images}} &
      \multicolumn{6}{|c|}{\bf{Simulation B-mode images}} &
      \multicolumn{4}{|c|}{\bf{Super Resolution images}} \\
      \hline 
        \hline
     & \bf{$\Delta$L} & \bf{$\Delta$D} & \bf{$I_c/I_u$} & \bf{RSL} & \bf{LSL} & \bf{MLW}  & \bf{$\Delta$L} & \bf{$\Delta$D} & \bf{$I_c/I_u$} & \bf{RSL} & \bf{LSL} & \bf{MLW} & \bf{$\Delta$L} & \bf{$\Delta$D} & \bf{Counts} & \bf{Counts} \\
     & \bf{($\mu$m)} & \bf{($\mu$m)} &  & \bf{(dB)} & \bf{(dB)} & \bf{($\mu$m)} & \bf{($\mu$m)} & \bf{($\mu$m)} &  & \bf{(dB)} & \bf{(dB)} & \bf{($\mu$m)} & \bf{($\mu$m)} & \bf{($\mu$m)} & & \bf{ratio} \\
    \hline
    \bf{Corrected} & 50 & 300 & 1.6 & 7 & 9 & 2533 & 62.5 & 210 & 2.76 & 12.3 & 21.7 & 2060 & 80 & 200 & 1040 & 1.48 \\
    \hline
    \bf{Uncorrected} & 990 & 1790 & - & 3.7 & 10.5 & 4454 & 625 & 620 & - & 13.3 & 12.2 & 2094 & 480 & 2000 & 700 & -  \\
    \hline
    \bf{Tube} & - & - & - & 16.1 & 14 & 2011 & - & - & - & 12.2 & 12.2 & 1630 & - & - & 1001 & - \\
    \hline
  \end{tabular}
  \end{scriptsize}
  \label{summary}
 
\end{table*}

The improvements from experimental phase aberration correction were compared with the  best-case scenario from the simulations. These simulated plots (Fig.~\ref{fig:beamplot}B), show the B-mode beamplots from the same three scenarios, namely for the corrected, uncorrected and microtube only emissions. The registration error laterally, is 62.5~$\mu$m for the corrected case and 625~$\mu$m for the uncorrected case, an improvement of 562.5~$\mu$m.

Axially, the registration error is 210~$\mu$m for the corrected B-mode image, and 620~$\mu$m for the uncorrected B-mode, an improvement of 410~$\mu$m. Additionally, the amplitude increased by a factor of 2.76 for the corrected B-mode when compared to the uncorrected counterpart.
 Fig.~\ref{fig:beamplot}B, shows that the shape of the tube without the skull is as expected for a point target that is perfectly centered. The uncorrected emission through the skull degrades the shape of the side lobes and causes a noticeable lateral error due to aberration effects from the skull. After applying the correction derived from the one-way distorted profile, one could expect an almost full restoration of the shape and position of the beamplot. However, that is not the case even in this gold standard setting. The width of the main lobe for the corrected emission is larger by 430~$\mu$m compared to the microtube, and the size and symmetry of the side lobes still deviates considerably, with the left side lobe at 21.7~dB lower than the main lobe and the right one at 12.3~dB lower. This implies that there are other, image-degrading effects caused by the skull, that cannot be reversed by means of phase correction only, namely amplitude distortion from attenuation and diffraction, and multiple reverberation in the skull.  

\begin{figure}[ht]
\setlength{\unitlength}{0.5\textwidth}
\begin{picture}(0.9,0.5)(0,0)
\put(0.0,0.0){\includegraphics[trim= 10 0 0 10,clip,height=0.49\unitlength]{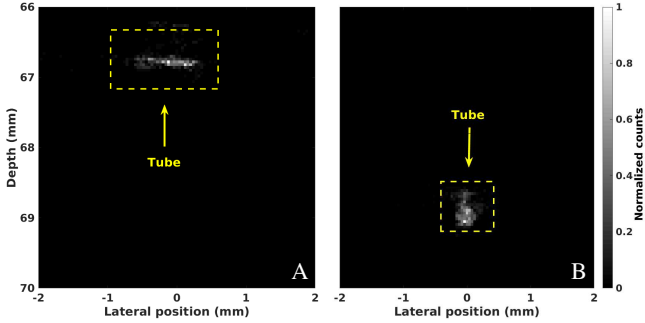}}
\color{white}
\put(0.42,0.07){\small A}
\put(0.83,0.07){\small B}
\end{picture}
\caption{Super resolved images for the (A) uncorrected and (B) corrected emissions through the skull. Scale in both cases is normalized number of bubble counts.}
\label{fig:SR}
\end{figure}

 The super resolved images produced from the out-of-plane experiment are shown in Fig.~\ref{fig:SR} for the conventional uncorrected emission in (A), and the corrected emission in (B), both through the skull. While qualitatively assessing the images, an evident degradation in the shape of the microtube is observed in (A), especially in the lateral direction. The shape after the application of the correction appears substantially improved and rounder in (B). Sensitivity in bubble detection for the corrected and uncorrected emissions through the skull, was calculated as the sum of bubble counts among all the accumulated frames (1000 used in this case). Specifically, in the uncorrected case 700 bubble events were detected compared to 1040 bubble events in the corrected case, which amounts to an improvement in sensitivity of a factor of 1.48, defined as the ratio of counts, for the corrected case. 

\begin{figure}[ht]
\setlength{\unitlength}{0.5\textwidth}
\begin{picture}(0.9,0.45)(0,0)
\put(0.0,0.0){\includegraphics[trim= 10 10 20 0,clip,height=0.5\unitlength]{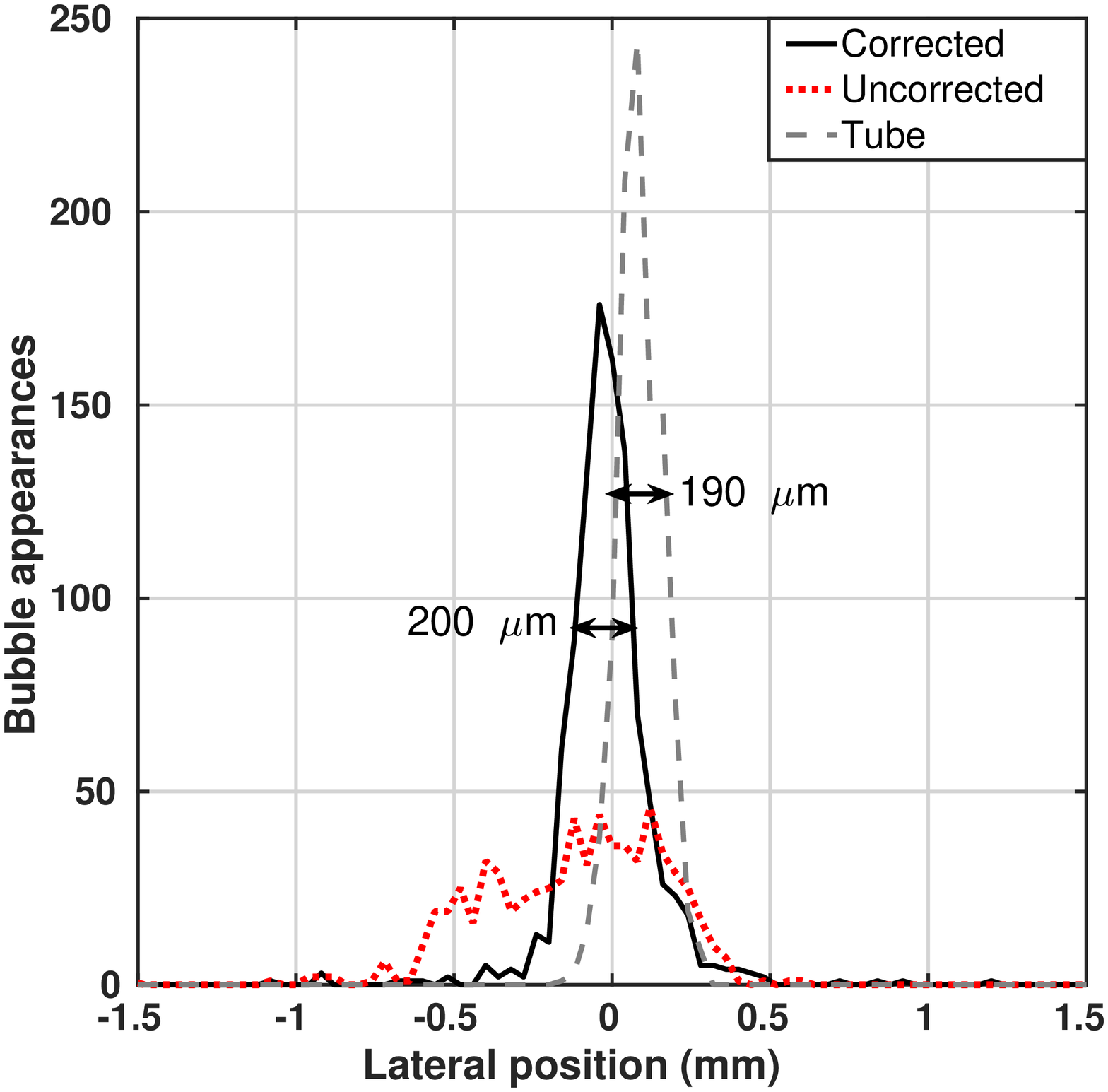}}
\put(0.51,0.0){\includegraphics[trim= 60 10 30 0,clip,height=0.5\unitlength]{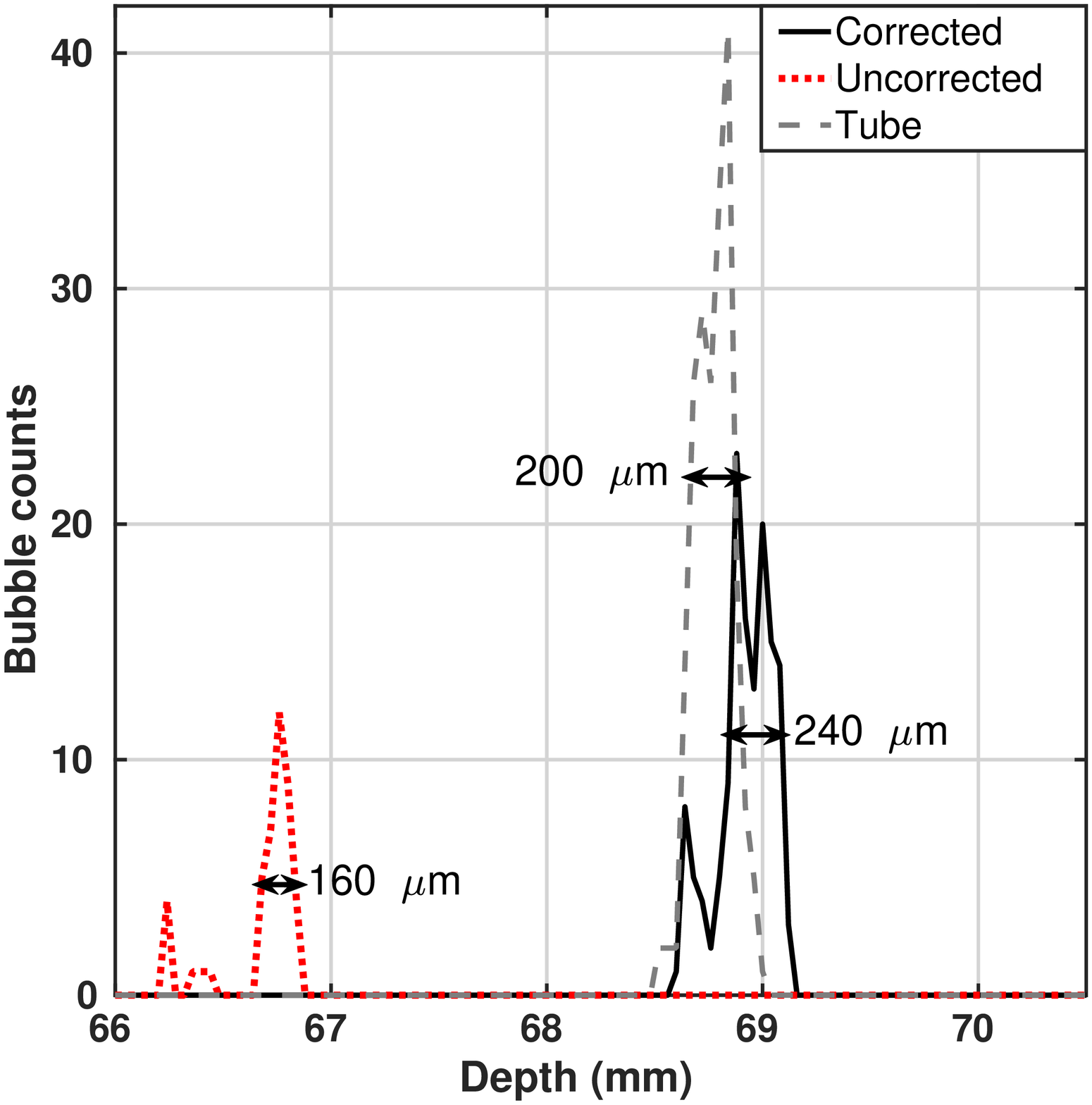}}
\put(0.43,0.06){\small A}
\put(0.91,0.06){\small B}
\end{picture}
\caption{Sum of bubble counts as a function of (A) lateral position and (B) depth, for the uncorrected and corrected super resolved images. The tube profiles are also provided for reference and registration error estimation ground truth. Sensitivity improves by a factor of 1.48 for the corrected case.} 
\label{fig:SRprof}
\end{figure}

To quantify the detection accuracy of the microtube in both directions, the sum of the bubble counts as a function of lateral (A) and axial (B) position are shown in Fig.~\ref{fig:SRprof}. The corresponding microtube only profiles are included for reference.  In Fig.~\ref{fig:SRprof}A, the lateral profiles have a registration error of 80~$\mu$m for the corrected super-resolved image, while for the uncorrected images the error is 480~$\mu$m, as estimated by the center of mass of the bubble positions. Size estimations of the detected microtube are also made by measuring the full width at half maximum (FWHM) for the curves. The corrected emission curve laterally estimates the size of the microtube at 200~$\mu$m, while in the uncorrected case, the size is estimated at 1~mm. Note that the actual size of the microtube is 208~$\mu$m.

In Fig.~\ref{fig:SRprof}B, the registration error in depth is 200~$\mu$m for the corrected super-resolved image and 2000~$\mu$m for the uncorrected counterpart, an improvement of 1800~$\mu$m. The corrected emission curve as a function of depth, estimates the FWHM axial size of the microtube at 240~$\mu$m, while in the uncorrected case the equivalent size is estimated at 160~$\mu$m. The deviation from the actual size in the axial direction is in the $\lambda$/20 range for the corrected emission profile.

These registration error estimations, for both the B-mode and super resolved images, in both directions, as well as beamplot characteristics are summarized in Table~\ref{summary}, where $\Delta$L denotes the lateral registration error, $\Delta$D for the registration error in depth, $I_c/I_u$ is the amplitude improvement ratio, RSL is the right side lobe amplitude, LSL the left side lobe amplitude and MLW stands for the main lobe width. Main lobe width estimations were performed at 6~dB of intensity. The estimates made from the B-mode images and the super-resolved images are consistent, with a maximum deviation of $\lambda$/20 between them.

\subsection{In-plane transcranial imaging} \label{in_results}
\begin{figure*}[ht]
\setlength{\unitlength}{0.95\textwidth}
\begin{picture}(0.5,0.4)(0,0)
\put(0.12,0.0){\includegraphics[trim= 0 0 0 0,clip,width=0.75\textwidth]{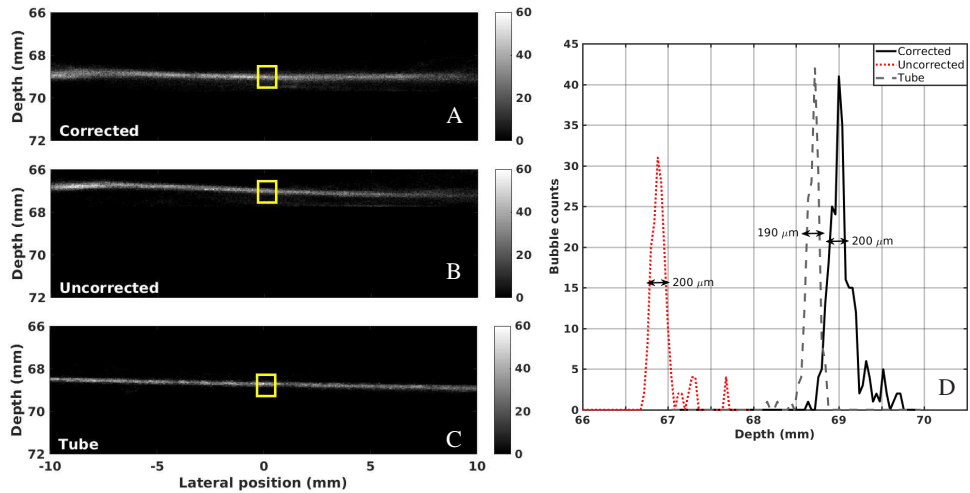}}
\put(0.85,0.09){\small D}
\color{white}
\put(0.47,0.3){\small A}
\put(0.47,0.18){\small B}
\put(0.47,0.05){\small C}
\end{picture}
\caption{Super resolved images for (A) the corrected emission through the skull, (B) the uncorrected emission through the skull and (C) the microtube without the skull. Scales are in number of bubble counts. (D) Laterally averaged counts as a function of depth for the corrected and uncorrected images with the skull and the microtube only as reference. Solid boxes show the location from where plot (D) was generated. }
\label{fig:inplane}
\end{figure*}

The super resolved images of the in-plane corrected and uncorrected emissions through the skull as well as the microtube without the skull are shown in Fig.~\ref{fig:inplane}, labeled as (A)-(C), respectively. For the uncorrected super-resolved image in (B), when compared to the microtube only super-resolved image in (C), a deviation from the true shape of the tube as well as an offset in depth can be qualitatively observed. However, even without applying a correction, the whole length of the microtube can still be resolved and observed, albeit with a large error in position. Both the shape and the offset in depth are improved after the application of the correction, as shown in Fig.~\ref{fig:inplane}A. For the corrected emission 79965 bubble appearance events are detected, which compared to 61198 events in the uncorrected emission reveal an improvement of 31$\%$ in sensitivity in bubble detection. As a reference, the total bubble counts for the tube without the skull, which were at a lower pulse pressure, are 53427.

Fig.~\ref{fig:inplane}D, shows the bubble sum profiles as a function of depth for a given central lateral position, which is highlighted in Fig.~\ref{fig:inplane}A -~\ref{fig:inplane}C with boxes. The profile of the microtube without the skull is provided as an actual position reference.
Axial size estimates of the detected microtube for all three aforementioned cases are given by the FWHM of each curve. In both uncorrected and corrected emissions through the skull, the axial size of the microtube is estimated at 200$\mu$m. The registration error between the true position of the microtube, the detected position for the uncorrected image, and the corrected image in depth was estimated  and then averaged for each lateral position spanning across the central 8~mm region of the tube. This yielded an estimate of 284$\pm$58~$\mu$m for the corrected emission and 1885$\pm$196~$\mu$m for the uncorrected emission.  

\section{Discussion} \label{discuss}
We have shown that super resolution imaging through a human skull specimen of out-of-plane and in-plane microtubes with a size of 208~$\mu$m ($\lambda$/3) is feasible at 2.5~MHz, even without phase correction. In the uncorrected out-of-plane emission case, the shape of the microtube was degraded laterally and the accuracy of the size of the microtube was estimated at 1~mm, while its central position was approximated at 480~$\mu$m off-center. Nevertheless, there was a sufficient number of microbubble appearances to indicate the presence and approximate position of the microtube. Further phase correcting reduced the lateral registration error to 80~$\mu$m as measured in the super resolved images. In the case of imaging the in-plane target through the skull, an adequate amount of microbubbles was detected to fully populate the vessel-like structure even without phase correcting. This however, may not always be the case, in an {\it in vivo} setting where attenuating tissue is present. Scattering from the tissue is expected to further decrease detectability of the microbubbles and may increase the need for correction. Moreover, the lateral degradation of the tube shape as seen in the out-of-plane case is not evident in the in-plane configuration due to the direction of the tube with respect to the imaging plane. Even when targeting a single focal position laterally, multiple neighbouring microbubble events are still detected along the lateral axis.

A single phase aberration correction profile was determined for the center of the field, and it was then applied to the entire field of view. This was performed under the assumption that a local correction is still effective across a lateral field of 30~mm. A more sophisticated version of the correction process would determine a specific correction for each imaging position, laterally and in depth.
Nevertheless the imaging results did not show a noticeable decrease in bubble detection events, resolution, or accuracy as a function of lateral position, suggesting that the small distance assumption was valid.  

Fullwave simulations of transcranial propagation, which provided a theoretical best-case scenario, showed improvements in both registration error and amplitude. The lateral registration error was 62.5~$\mu$m for the corrected image, and there was an improvement of almost $\lambda$ from the uncorrected image, as well as a significant amplitude correction of a factor of 2.74. These simulation results therefore suggest that there may be room for improvement in the experimental phase aberration correction implementation. However the simulations also show that the width and shape of the main lobe and side lobes are not fully restored with phase aberration correction. This phenomenon was also observed experimentally and it is consistent with the interpretation that amplitude aberration and multiple reverberation also play a significant role in image quality.  Except for phase distortion, the skull also reduces the amplitude of the initial pulse, which may be partially compensated with with amplitude correction techniques.
 
 A limitation of the method as it was implemented, is that it is challenging to extract a correction from an aligned point target {\it in vivo}, since it is impossible to find such a target occurring naturally. This limitation has been previously addressed in focused ultrasound therapy, where pre-calculated simulation derived corrections for a specific registered location of the skull can be used~\cite{Almquist2016}. These corrections can be produced as described in subsection~\ref{sims} and, provided the CT scan used is accurately registered to the actual skull, they can be employed to phase correct a given region of interest, as previously performed in therapy transcranial ultrasound~\cite{cttherapy}.
 
 \section{Conclusion}
 In summary and conclusion, transcranial super resolution imaging through a human skull is feasible at a frequency of 2.5~MHz, even without applying a phase correction. With the placement of the transducer in the temporal skull window when combined with phase correction, images of vessel-mimicking microtubes of a diameter of 208~$\mu$m with a lateral registration error as low as 80~$\mu$m were produced. The correction in the out-of-plane configuration thus reduced the error to the $\lambda$/8 range and increased sensitivity by a factor of 1.48 in the super-resolution images, as compared to the uncorrected emission. In the in-plane configuration the microtube was detectable with and without correction, however the correction increased sensitivity by a factor of 1.31 as compared to the uncorrected counterpart. This method has the potential to significantly improve the resolution and accuracy necessary for brain vascular mapping scenarios.
 
\ifCLASSOPTIONcaptionsoff
  \newpage
\fi

\bibliographystyle{IEEEtran}
\bibliography{IEEEabrv,ref}

\end{document}